\documentstyle[12pt]{article}

\newcommand{\bce}{\begin{center}}
\newcommand{\ece}{\end{center}}
\newcommand{\beq}{\begin{equation}}
\newcommand{\eeq}{\end{equation}}
\newcommand{\bea}{\vspace{0.25cm}\begin{eqnarray}}
\newcommand{\eea}{\end{eqnarray}}

\newcommand{\ba}{\begin{array}}
\newcommand{\ea}{\end{array}}

\def\lsim{\mathrel{\rlap{\lower4pt\hbox{\hskip1pt$\sim$}}
    \raise1pt\hbox{$<$}}}         
\def\gsim{\mathrel{\rlap{\lower4pt\hbox{\hskip1pt$\sim$}}
    \raise1pt\hbox{$>$}}}         

\def\Pom{{\bf I\!P}}

\def\lsim{\mathrel{\rlap{\lower4pt\hbox{\hskip1pt$\sim$}}
    \raise1pt\hbox{$<$}}}         
\def\gsim{\mathrel{\rlap{\lower4pt\hbox{\hskip1pt$\sim$}}
    \raise1pt\hbox{$>$}}}         

\def\Pom{{\bf I\!P}}

  \advance\oddsidemargin  by -0.9in
  \advance\evensidemargin by -0.9in
  \advance\topmargin      by -0.20in
\makeindex
\begin{document}

\phantom{.}\hspace{10.5cm}{\large \bf 24 March 2000}
\vspace{1.5cm}\\
\begin{center}
{\Large \bf
The wave function
of 2S radially excited vector mesons from data
for diffraction slope} \\
\vspace*{1.5cm} 
{\large \bf J.~Nemchik}
\vspace*{0.5cm} \\
{\it Institute of Experimental Physics, Slovak Academy of Sciences, \\
Watsonova 47, 04353 Ko\v sice, Slovakia} \\
\vspace*{3.5cm}
\begin{minipage}[h]{13cm}
\centerline{\Large \bf
Abstract }
\vspace*{1.5cm}

In the color dipole gBFKL dynamics,
we predict a strikingly different 
$Q^{2}$ and energy dependence of the diffraction slope
for the elastic production of ground state $V(1S)$ and 
radially excited $V'(2S)$ light
vector mesons.
The color dipole model predictions for the 
diffraction
slope for $\rho^{0}$ and $\phi^{0}$ production
are in a good
agreement with the data from the 
fixed target and collider HERA experiments.
We present how a different form of anomalous
energy- and $Q^{2}$ dependence of the diffraction
slope for $V'(2S)$ production leads to a different
position of the node in radial wave function
and discuss a possibility how to determine this
position from the fixed target and HERA data.

\end{minipage}
\end{center}
\pagebreak
\setlength{\baselineskip}{0.55cm}
%

\section{Introduction}

Diffractive photo- and electroroduction of 
vector mesons 
%
%
\beq
\gamma^{*}p \rightarrow Vp\, ,\,\, (V = \rho^{0}, \Phi^{0},
\omega^{0}, J/\Psi, \Upsilon ...)
\label{eq:1}
\eeq
%
%
is presently intensively studied
at HERA and represent a good cross check
to test the ideas implemented into various 
theoretical models 
\cite{DL,KZ91,Ryskin,KNNZ93,KNNZ94,NNZscan,Brodsky,Forshaw,GLM}
within the framework of the perturbative QCD (pQCD). 
Morevever, the high statistics data 
at HERA
during a several last years allows also
to study diffractive electroproduction
of radially excited $V'(2S)$ vector mesons,
which are known to have a node
(the node effect 
\cite{KZ91,KNNZ93,NNZanom,NNPZ97,NNPZZ98}) in the radial
wave function leading to pecularities in
investigation of various aspects 
of their diffractive production. 
In this paper we demonstrate
further salient features of the node effect
in conjunction with the gBFKL phenomenology of the
diffraction slope 
\cite{NZZslope,NZZspectrum,NNPZZ98}
leading to an anomalous
energy and $Q^{2}$ dependence of the diffraction cone.

The details of the gBFKL phenomenology of diffractive
electroproduction of vector mesons has been presented
in the paper \cite{NNPZ97} and will not be
repeated here. The same concerns to the color dipole phenomenology
of the diffraction slope for photo- and electroproduction
of heavy vector mesons developed in the paper \cite{NNPZZ98}.
We start with the principal result coming from
the analysis of the diffractive production of light
\cite{NNZscan,NNPZ97} and heavy
\cite{NNPZZ98} vector mesons at $t=0$ within
the gBFKL phenomenology and leading to the conclusion
that the
$1S$ vector meson production amplitude probes the color dipole cross
section
(and the dipole diffraction slope as well)
at the dipole size $r\sim r_{S}$
({\it scanning phenomenon} \cite{NNN92,KNNZ93,KNNZ94,NNZscan}),
where the scanning radius can be expressed through the scale
parameter $A$, photon virtuality $Q^{2}$ and
vector meson mass $m_{V}$ :
%
\beq
r_{S} \approx {A \over \sqrt{m_{V}^{2}+Q^{2}}}\, .
\label{eq:2}
\eeq
%
Scanning phenomenon allows to study the transition
between the perturbative (hard)
and nonperturbative (soft) regimes.
Changing $Q^{2}$ and the mass of the produced
vector meson, one can
probe the dipole cross section $\sigma(\xi,r)$,
and the dipole diffraction slope $B(\xi,r)$
in a very broad range of the dipole
sizes, $r$.

Radially excited $V'(2S)$ vector mesons can extend
an additional
information on the dipole cross section and dipole
diffraction slope.
The presence of the node in the
$2S$ radial
wave function leads to the node effect
(a strong cancellation
of the dipole size contributions to the production amplitude
from the region above and below the node position,
$r_{n}$, in the $2S$ radial wave function \cite{KZ91,NNN92,NNZanom,NNPZ97}).
For this reason, the amplitudes for the
electroproduction of the $1S$ and $2S$ vector mesons probe $\sigma(\xi,r)$
and $B(\xi,r)$ in a different way.
The onset of the node effect depends
on vector meson mass.
The node effect has been found to be a strong in
electroproduction of radially excited
light vector mesons ($\rho^{0}$, $\Phi^{0}$, $\omega^{0}$)
\cite{NNPZ97} leading
to an anomalous $Q^{2}$ and energy dependence of the production
cross section.
However, node effect is much weaker for the electroproduction of $2S$ heavy
vector mesons ($J/\Psi$, $\Upsilon$, ...)
For production of charmonia it
leads to a slightly different $Q^{2}$ and
energy dependence of the production cross section
for $\Psi'$ vs. $J/\Psi$ 
and to a counterintuitive inequality $B(\Psi') < B(J/\Psi)$
\cite{NNPZZ98}.

For $\Upsilon'$ production, the node
effect is negligible small and gives approximately the
same $Q^{2}$ and energy behaviour of the production cross section
and practically the same diffraction slope at $t=0$ for
$\Upsilon$ and $\Upsilon'$ production \cite{NNPZZ98}.
Therefore, it is very important to explore farther the salient features
of the node effect with conjunction with the emerging gBFKL
phenomenology of the diffraction slope especially in production of
$V'(2S)$ light vector mesons where the node effect is
expected to be very strong.

Two main reasons affect the cancellation pattern
in the diffraction slope for $2S$ state.
The first reason is connected with the $Q^{2}$ behaviour
of the scanning radius $r_{S}$ (see (\ref{eq:2}));
for the electroproduction of $V'(2S)$ light vector mesons
at moderate $Q^{2}$ when the scanning radius
$r_{S}$ is close to $r_{n}$,
due to $\sim r^{2}$ behaviour of $B(\xi,r)$ \cite{NZZslope}
even a slight variation of
$r_{S}$ with $Q^{2}$ strongly changes the cancellation
pattern  and leads to an anomalous
$Q^{2}$ dependence
of the forward diffraction slope, $B(t=0)$
\cite{NNPZZ98}.
The second reason is due to
different energy dependence
of $\sigma(\xi,r)$
at different dipole sizes $r$ coming from
the gBFKL dynamics
leading also to an
anomalous energy dependence of $B(t=0)$ for the $V'(2S)$ production.

The effects mentioned above are sensitive to the
form of the dipole cross section and the dipole diffraction slope.
In Ref.~\cite{NNPZdipole} (\cite{SlopeDipole})
we presented the first direct determination
of the color dipole cross section (color dipole diffraction slope)
from the data on
the photo- and electroproduction of $V(1S)$ vector mesons.
So extracted dipole cross section 
(dipole diffraction slope) is in a good agreement with
the dipole cross section (dipole diffraction slope)
obtained from gBFKL analysis
\cite{NZHera,NNZscan} (\cite{NZZslope,NZZspectrum,NNPZZ98}).
This fact confirms a very reasonable choice
of the nonperturbative component of the dipole cross section
(dipole diffraction slope)
corresponding to a soft nonperturbative mechanism contribution
to the scattering amplitude.

Due to a large value of the scale parameter in (\ref{eq:2}),
the large-distance
contributions to the production amplitude
from the semiperturbative and nonperturbative
region of color dipoles $r\gsim R_{c}$ becomes substantial
($R_{c}\sim 0.27 fm$ is gluon correlation radius introduced in
\cite{NZ94,NZZ94}) .
Only the virtual $\rho^{0}$ and $\phi^{0}$ photoproduction
at $Q^{2}\gsim 100$\,GeV$^{2}$ can be treated as a purely
perturbative process, when the production
amplitude is dominantly contributed from the perturbative
region, $r\lsim R_{c}$.

In the present paper
we concentrate on the production of $V'(2S)$ radially
excited light vector mesons, where the node in the
radial wave function in conjunction with the
subasymptotic energy dependence of $B(\xi,r)$
leads to a
strikingly different $Q^{2}$ and energy
dependence of the diffraction slope
for the production of $V'(2S)$ vs.
$V(1S)$ vector mesons.
We also study how
the position of the node in the radial wave function
for $V'(2S)$ vector mesons can be
extracted from the data.
We present an exact prescription how the experimental
measurement of the $Q^{2}$ and energy
dependence of the diffraction slope for
$V'(2S)$ production could
distinguish between the undercompensation
and overcompensation scenarios of the $2S$ production amplitude
(see Section 4).
The explicit form of that $Q^{2}$- and energy behaviour
of the diffraction slope is connected
with the position of the node in radial wave function
for $V'(2S)$ vector mesons.
This paper is organized as follows. In section 2 we present 
a very short review of the 
the color dipole phenomenology
of the diffractive photo- and electroproduction of vector mesons
including some needful results from 
the gBFKL phenomenology of the diffraction slope.
Section 3 contains the model predictions
for $Q^{2}$ and energy dependence of the forward diffraction slope 
for the $\rho^{0}$ and $\phi^{0}$ real and virtual
electroproduction.
We predict a substantial growth of the diffraction slope
with energy in a good agreement
with the low energy data and
the data from the HERA collider experiments.
The subject of section 4 concerns to
the anomalous $Q^{2}$ and energy dependence of diffraction slope
for electroproduction of $2S$ radially excited light vector mesons.
The summary
and conclusions are presented in section 5.



\section{Basic formulas from the color dipole phenomenology
         of vector meson production and the diffraction slope}

In the mixed $({\bf{r}},z)$ representation,
the high energy meson is considered as
a system of color dipole described by
the distribution
of the transverse separation ${\bf{r}}$ of the quark and
antiquark given by the $q\bar{q}$ wave function,
$\Psi({\bf{r}},z)$, where $z$ is
the fraction of meson's lightcone momentum
carried by a quark.
The Fock state expansion for the
relativistic meson starts
with the $q\bar{q}$ state and
the higher Fock states $q\bar{q}g...$
become very important at high energy $\nu$.
The interaction of the relativistic
color dipole of the dipole moment, ${\bf{r}}$, with the
target nucleon is quantified by the energy dependent color
dipole cross section, $\sigma(\xi,r)$,
satisfying
the gBFKL equation
\cite{NZ94,NZZ94} for the energy evolution.
This reflects the fact that 
in the leading-log ${1\over x}$ approximation the
effect of higher Fock states can be
reabsorbed into the energy dependence
of $\sigma(\xi,r)$.
The dipole cross section is flavor
independent and represents the universal
function of $r$ which describes
various diffractive processes in unified form.
At high energy, when the transverse separation, ${\bf{r}}$,
of the quark and antiquark is frozen during the interaction
process, 
the scattering
matrix describing the $q\bar{q}$-nucleon interaction
becomes diagonal
in the mixed $({\bf{r}},z)$-representation ($z$ is known also as
the Sudakov light cone variable). 
This diagonalization property is held even 
when the dipole size, ${\bf{r}}$, is large,
i.e. beyond the perturbative region of short distances.

Following an advantage of
the $({\bf{r}},z)$-diagonalization of the
$q\bar{q}-N$ scattering matrix, the
imaginary part of the production
amplitude for the real (virtual) photoproduction
of vector mesons
with the momentum transfer ${\bf{q}}$ can be represented in the 
factorized form
\beq
{\rm Im}{\cal M}(\gamma^{*}\rightarrow V,\xi,Q^{2},{\bf{q}})=
\langle V |\sigma(\xi,r,z,{\bf{q}})|\gamma^{*}\rangle=
\int\limits_{0}^{1} dz\int d^{2}{\bf{r}}\sigma(\xi,r,z,{\bf{q}})
\Psi_{V}^{*}({\bf{r}},z)\Psi_{\gamma^{*}}({\bf{r}},z)\,
\label{eq:3}
\eeq
whose normalization is
$
\left.{d\sigma/ dt}\right|_{t=0}={|{\cal M}|^{2}/ 16\pi}.
$
In Eq.~(\ref{eq:3}), 
$\Psi_{\gamma^{*}}({\bf{r}},z)$ and
$\Psi_{V}({\bf{r}},z)$ represent the
probability amplitudes
to find the color dipole of size, $r$,
in the photon and quarkonium (vector meson), respectively.
The color dipole distribution in (virtual) photons was
derived in \cite{NZ91,NZ94}.
$\sigma(\xi,r,z,{\bf{q}})$ 
is the dipole scattering matrix for $q\bar{q}-N$ interaction. 
and
represents the above mentioned color dipole cross section
for ${\bf{q}}=0$.
At small ${\bf{q}}$ considered in this paper,
one can safely neglect
the $z$-dependence of $\sigma(\xi,r,z,{\bf{q}})$ 
for light and heavy vector meson production
and set $z=\frac{1}{2}$.
This follows partially from the analysis within double gluon 
exchange approximation
\cite{NZ91} leading to a slow $z$ dependence of 
the dipole cross section.

The energy dependence of the dipole cross section is quantified
in terms of the dimensionless
rapidity, $\xi=\log{1\over x_{eff}}$, where
$x_{eff}$ is the effective value of the Bjorken variable
%
%
\beq
x_{eff} =
\frac {Q^{2}+m_{V}^{2}}{Q^{2}+W^{2}} \approx
 \frac{m_{V}^{2}+Q^{2}}{2\nu m_{p}}\, ,
\label{eq:4}
\eeq
%
%
where $m_{p}$ is the proton mass.
Hereafter, we will write the energy dependence of the dipole
cross section in both variables,
either in $\xi$ or in $x_{eff}$. 

The production amplitudes for the
transversely (T) and the longitudinally (L) polarized vector mesons
with the momentum transfer, $\bf{q}$,
can be written in more explicit form \cite{NNZscan,NNPZZ98}
%
%
\bea
{\rm Im}{\cal M}_{T}(x_{eff},Q^{2},{\bf{q}})=
{N_{c}C_{V}\sqrt{4\pi\alpha_{em}} \over (2\pi)^{2}}
\cdot~~~~~~~~~~~~~~~~~~~~~~~~~~~~~~~~~
\nonumber \\
\cdot \int d^{2}{\bf{r}} \sigma(x_{eff},r,{\bf{q}})
\int_{0}^{1}{dz \over z(1-z)}\left\{
m_{q}^{2}
K_{0}(\varepsilon r)
\phi(r,z)-
[z^{2}+(1-z)^{2}]\varepsilon K_{1}(\varepsilon r)\partial_{r}
\phi(r,z)\right\}\nonumber \\
 =
{1 \over (m_{V}^{2}+Q^{2})^{2}}
\int {dr^{2} \over r^{2}} {\sigma(x_{eff},r,{\bf{q}}) \over r^{2}}
W_{T}(Q^{2},r^{2})
\label{eq:5}
\eea
%
%
\bea
{\rm Im}{\cal M}_{L}(x_{eff},Q^{2},{\bf{q}})=
{N_{c}C_{V}\sqrt{4\pi\alpha_{em}} \over (2\pi)^{2}}
{2\sqrt{Q^{2}} \over m_{V}}
\cdot~~~~~~~~~~~~~~~~~~~~~~~~~~~~~~~~~
 \nonumber \\
\cdot \int d^{2}{\bf{r}} \sigma(x_{eff},r,{\bf{q}})
\int_{0}^{1}dz \left\{
[m_{q}^{2}+z(1-z)m_{V}^{2}]
K_{0}(\varepsilon r)
\phi(r,z)-
\varepsilon K_{1}(\varepsilon r)\partial_{r}
\phi(r,z)\right\} \nonumber \\
 =
{1 \over (m_{V}^{2}+Q^{2})^{2}}
{2\sqrt{Q^{2}} \over m_{V}}
\int {dr^{2} \over r^{2}} {\sigma(x_{eff},r,{\bf{q}}) \over r^{2}}
W_{L}(Q^{2},r^{2})
\label{eq:6}
\eea
%
%
where
%
%
\beq
\varepsilon^{2} = m_{q}^{2}+z(1-z)Q^{2}\,,
\label{eq:7}
\eeq
%
%
$\alpha_{em}$ is the fine structure 
constant, $N_{c}=3$ is the number of colors,
$C_{V}={1\over \sqrt{2}},\,{1\over 3\sqrt{2}},\,{1\over 3},\,
{2\over 3},\,{1\over 3}~~$ for 
$\rho^{0},\,\omega^{0},\,\phi^{0},\, J/\Psi, \Upsilon$ production,
respectively and
$K_{0,1}(x)$ are the modified Bessel functions.
The detailed discussion and parameterization 
of the lightcone radial wave function $\phi(r,z)$
of the $q\bar{q}$ Fock state of the vector meson
is given in \cite{NNPZ97}.

The terms $\propto \epsilon K_{1}(\epsilon r)\partial_{r}\phi({\bf r},z)$
for $T$ polarization
and $\propto K_{0}(\epsilon r)\partial_{r}^{2}\Phi({\bf r},z)$
for $L$ polarization
in the integrands of
(\ref{eq:5}) and (\ref{eq:6}) represent
the relativistic corrections
which become important
at large $Q^{2}$ and for
the production of light vector mesons.
For the production of heavy quarkonia,
the nonrelativistic approximation can be used
with a rather high accuracy \cite{KZ91}.

The weight functions,
$W_{T}(Q^{2},r^{2})$ and
$W_{L}(Q^{2},r^{2})$, introduced in (\ref{eq:5}) and (\ref{eq:6})
have a smooth $Q^{2}$ behaviour \cite{NNZscan} and are very
convenient for the analysis of the scanning phenomenon.
They are
sharply peaked at $r\approx A_{T,L}/\sqrt{Q^{2}+m_{V}^{2}}$.
At small $Q^{2}$ the values of the scale parameter $A_{T,L}$ are 
close to $A \sim 6$, which follows from $r_{S}=3/\varepsilon$ with
the nonrelativistic choice $z=\frac{1}{2}$. 
In general, $A_{T,L} \geq 6$
and increases slowly with $Q^2$ \cite{NNZscan}.
For heavy vector meson production, the scale parameters $A_{T,L}
\sim 6$ for $\Upsilon$ at $Q^{2}\le 100$\,GeV$^{2}$ and
$A_{T,L}\sim 6$ at $Q^{2}=0$ and
$A_{T,L}\sim 7$ at $Q^{2}=100\,$GeV$^{2}$ for $J/\Psi$.
For this reason, 
the heavy vector mesons can be treated nonrelativistically,
except for small relativistic corrections for the electroproduction
of charmonia at very large $Q^{2}\sim 100$\,GeV$^{2}$. 
Not so for the light vector mesons where the relativistic
corrections play an important role especially at large
$Q^{2}\gg m_{V}^{2}$, and lead to
$Q^{2}$ dependence of $A_{L,T}$ coming from
the large-size asymmetric $q\bar{q}$ configurations:
$A_{L}(\rho^0;Q^{2}=0)\approx 6.5,~
A_{L}(\rho^0;Q^2 = 100\,{\rm GeV}^2) \approx 10,~
A_{T}(\rho^0;Q^{2}=0) \approx 7,~
A_{T}(\rho^0,Q^2 = 100\,{\rm GeV}^2)\approx 12$ \cite{NNZscan}.
Due to an extra factor $z(1-z)$ in the integrand of
(\ref{eq:6}) in comparison with (\ref{eq:5}),
the contribution from asymmetric $q\bar{q}$ 
configurations to the longitudinal
meson production is considerably smaller.

The integrands in 
Eqs.~(\ref{eq:5}) and
(\ref{eq:6}) contain the dipole cross section,
$\sigma(\xi,r,{\bf{q}})$.
As was mentioned,
due to a very slow onset of the pure perturbative region
(see Eq.~(\ref{eq:2})),
one can easily anticipate 
a contribution to the production amplitude
coming
from the semiperturbative and nonperturbative $r\gsim R_{c}$.
Following the simplest assumption about an additive property
of the perturbative and nonperturbative mechanism of interaction,
we can represent the contribution of the bare pomeron exchange
to $\sigma(\xi,r,{\bf{q}})$ as a sum
of the perturbative and nonperturbative component
\cite{NNPZ97,NNPZZ98}
%
%
\beq
\sigma(\xi,r,{\bf{q}}) = 
\sigma_{pt}(\xi,r,{\bf{q}})+\sigma_{npt}(\xi,r,{\bf{q}})\,,
\label{eq:8}
\eeq
%
%
with the parameterization of both components at small ${\bf{q}}$
\beq
\sigma_{pt,npt}(\xi,r,{\bf{q}})=\sigma_{pt,npt}(\xi,r,{\bf{q}}=0)
\exp\Bigl(-\frac{1}{2}
B_{pt,npt}(\xi,r){\bf{q^{2}}}\Bigr)\,.
\label{eq:2.9}
\eeq
Here $\sigma_{pt,npt}(\xi,r,{\bf{q}}=0)
= \sigma_{pt,npt}(\xi,r)$ represent the contribution
of the perturbative and nonperturbative mechanisms to the
$q\bar{q}$-nucleon interaction cross section, 
respectively, $B_{pt,npt}(\xi,r)$ are the
corresponding
diffraction slopes. 

A small real part of  production amplitudes can be taken
in the form \cite{GribMig}
%
%
\beq
{\rm Re}{\cal M}(\xi,r) =\frac{\pi}{2}\cdot\frac{\partial}
{\partial\xi} {\rm Im}{\cal M}(\xi,r)\,.
\label{eq:10}
\eeq
%
%

and can be easily included in the production amplitudes
(\ref{eq:5}),(\ref{eq:6})
using substitution
%
%
\beq
\sigma(x_{eff},r,{\bf q})\rightarrow
\biggl (1-i\frac{\pi}{2}\frac{\partial}{\partial~log~x_{eff}}
\biggr)
\sigma(x_{eff},r) = \biggl [1-i\alpha_{V}(x_{eff},r)
\biggr  ]\sigma(x_{eff},r,{\bf q})
\label{eq:11}   
\eeq
%
%

The formalism for calculation of $\sigma_{pt}(\xi,r)$
in the leading-log $s$ approximation was developed
in \cite{NZ91,NZ94,NZZ94}. 
The nonperturbative contribution, $\sigma_{npt}(\xi,r)$, 
to the dipole cross section was used in
Refs.~\cite{NZHera,NNZscan,NNPZ97,NNPZZ98} where
we assume that this soft nonperturbative component
of the pomeron is a simple Regge pole with
the intercept, $\Delta_{npt}=0$.
The particular form together with 
assumption of
the energy independent
$\sigma_{npt}(\xi=\xi_{0},r)=\sigma_{npt}(r)$
($\xi_{0}$ corresponds to boundary condition for the gBFKL
evolution, $\xi_{0}=\log{1/x_{0}}$, $x_{0}=0.03$)
allows one to successfully describe \cite{NZHera} the
proton structure function at very small $Q^{2}$,
the real photoabsorption \cite{NNZscan} and
diffractive real and virtual photoproduction of light
\cite{NNPZ97} and heavy \cite{NNPZZ98} vector mesons.
A larger contribution of the nonperturbative pomeron
exchange to $\sigma_{tot}(\gamma p)$ vs.
$\sigma_{tot}(\gamma^{*} p)$ can, for example, explain
a much slower
rise with energy
of the real photoabsorption cross section, 
$\sigma_{tot}(\gamma p)$, in comparison
with $F_{2}(x,Q^{2})\propto
\sigma_{tot}(\gamma^{*} p)$ observed at HERA \cite{H1sf,ZEUSsf}.
Besides, the reasonable form of this soft cross section, $\sigma_{npt}(r)$,
was confirmed in the process of the first determination of the dipole
cross section from the experimental data on vector meson
electroproduction \cite{NNPZdipole}. The so extracted dipole cross section
is in a good agreement with the dipole cross section obtained
from the gBFKL dynamics \cite{NNZscan,NZHera}.
Thus, this nonperturbative component of the pomeron
exchange plays a dominant
role at low NMC energies
in the production of the light vector mesons, where the
scanning radius, $r_{S}$ (\ref{eq:2}), is large.
However, the perturbative component of the pomeron become
more important with the rise of energy also in the nonperturbative
region of the dipole sizes. 

Now we present the basic aspects of the diffraction
slope coming
from the gBFKL phenomenology \cite{NZZslope,NNPZZ98}. 
As the result of the generalization of the factorization
formula (\ref{eq:3}) to the diffraction slope of the
reaction $\gamma^{*}p\rightarrow Vp$ one can write
%
%
\beq
B(\gamma^{*}\rightarrow V,\xi,Q^{2})
{\rm Im} {\cal M}(\gamma^{*}\rightarrow V,\xi,Q^{2},\vec{q}=0)=
\int\limits_{0}^{1} dz\int d^{2}\vec{r}\lambda(\xi,r)
\Psi_{V}^{*}(r,z)\Psi_{\gamma^{*}}(r,z)\,.
\label{eq:12}
\eeq
%
%
where
%
%
\beq
\lambda(\xi,r)=\int d^2\vec{b}~
\vec{b}\,^2~\Gamma(\xi,\vec{r,}\vec{b})\, .
\label{eq:13}
\eeq
%
%
Then the diffraction slope expressed through
the amplitude of elastic scattering of the color
dipole ${\rm Im}{\cal M}$
%
%
\beq
B(\xi,r)=\left.-
2{d \log {\rm Im}{\cal M}(\xi,r,{\bf q})/ dq^{2}}\right|_{q=0}
\label{eq:14}
\eeq
%
%
equals
%
%
\beq
B(\xi,r)= {1\over 2}\langle \vec{b}\,^{2}\rangle =
\lambda(\xi,r)/\sigma(\xi,r)\,.
\label{eq:15}
\eeq
%
%
The amplitude ${\rm Im}{\cal M}(\xi,r,{\bf q})$
in (\ref{eq:14}) 
within the impact-parameter representation
reads
%
%
\beq
{\rm Im} {\cal M}(\xi,r,\vec{q})=2\int d^{2}\vec{b}\,
\exp(-i\vec{q}\vec{b})\Gamma(\xi,\vec{r},\vec{b})\,,
\label{eq:16}
\eeq
%
%
where $\Gamma(\xi,r,{\bf b})$ is the profile function
and ${\bf b}$ is the impact parameter defined with
the respect to the center of the $q\bar{q}$ dipole.

The diffraction cone in the color dipole gBFKL approach
for production of vector mesons has been detaily studied
in \cite{NNPZZ98}. Here we only present the salient
feature of the color diffraction slope reflecting
the presence of the geometrical contribution from beam
dipole - $r^{2}/8$
and the contribution from the target proton size - $R_{N}^{2}/3$:
%
%
\beq
B(\xi,r)=
\frac{1}{8}r^{2}+\frac{1}{3}R_{N}^{2}+
2\alpha_{\Pom}'(\xi-\xi_{0}) + {\cal O}(R_{c}^{2})\, ,
\label{eq:17}
\eeq
%
%
where $R_{N}$ is the radius of the proton.
For electroproduction of light vector mesons the
scanning radius is larger than the correlation one
$r\gsim R_{c}$ even for $Q^{2}\lsim 50 GeV^{2}$
and one recovers a sort of
additive quark model, in which the uncorrelated gluonic clouds
build up around the beam and target quarks and antiquarks and
the term $2\alpha_{\Pom}'(\xi-\xi_{0})$
describe the familiar Regge growth of diffraction slope for
the quark-quark scattering.
The geometrical contribution to the diffraction
slope from the target proton size, ${1\over 3}R_{N}^{2}$,
persists for all the dipole sizes,
$r\gsim R_{c}$ and $r\lsim R_{c}$. The last term in (\ref{eq:17})
is also associated with the proton size and is negligibly small.

The soft pomeron and diffractive scattering of large color dipole has been
also detaily studied in the paper \cite{NNPZZ98}.
Here we assume the conventional Regge rise of the diffraction
slope for the soft pomeron,
%
%
\beq
B_{npt}(\xi,r)=\Delta B_{d}(r)+\Delta B_{N}+
2\alpha_{npt}^{'}(\xi-\xi_{0})\,,
\label{eq:18}
\eeq
%
%
where $\Delta B_{d}(r)$ and $\Delta B_{N}$ stand for the contribution
from the beam dipole and target nucleon size.
As a guidance we take the experimental
data on the pion-nucleon scattering
\cite{Schiz}, which suggest $\alpha'_{npt}=0.15$\,GeV$^{-2}$.
In (\ref{eq:18}) the proton size contribution
is
%
%
\beq
\Delta B_{N}={1\over 3}R_{N}^{2}\, ,
\label{eq:19}
\eeq
%
%
and
the beam dipole contribution has been proposed
to have a form
%
%
\beq
B_{d}(r) = {r^{2} \over 8}\cdot
{r^{2}+aR_{N}^{2} \over 3r^{2}+aR_{N}^{2}}\,,
\label{eq:20}
\eeq
%
%
where $a$ is a phenomenological parameter, $a\sim 1$.
We take $\Delta B_{N}=4.8\,{\rm GeV}^{-2}$.
Then the pion-nucleon diffraction slope is reproduced with
reasonable values of the parameter $a$ in the formula (\ref{eq:20}):
$a=0.9$ for $\alpha'_{npt}=0.15$\,GeV$^{-2}$ \cite{NNPZZ98}.

Following the simple geometrical properties
of the gBFKL diffraction slope, $B(\xi,r)$, (see Eq.~(\ref{eq:17})
and \cite{NZZslope}),
one can express its energy dependence through the energy
dependent effective Regge slope, $\alpha_{eff}'(\xi,r)$
%
%
\beq
B_{pt}(\xi,r) \approx \frac{1}{3}<R_{N}^{2}> + \frac{1}{8}r^{2}
+ 2\alpha_{eff}'(\xi,r)(\xi-\xi_{0}).
\label{eq:21}
\eeq
%
%
The effective Regge slope, $\alpha_{eff}'(\xi,r)$,
varies
with energy differently
at different size of the color dipole
\cite{NZZslope};
at fixed scanning radius and/or $Q^{2}+m_{V}^{2}$,
it decreases with energy.
At fixed rapidity $\xi$
and/or $x_{eff}$ (\ref{eq:4}),
$\alpha_{eff}'(\xi,r)$
rises with $r\lsim 1.5$\,fm.
At fixed energy, it is a flat function
of the scanning radius.
At the asymptotically large $\xi$ ($W$),
$\alpha_{eff}'(\xi,r)\rightarrow \alpha_{\Pom}'=0.072$\,GeV$^{-2}$.
At the lower and HERA energies, the subasymptotic
$\alpha_{eff}'(\xi,r)\sim (0.15-0.20)$\,GeV$^{-2}$ and is very
close to $\alpha_{soft}'$ known from the Regge phenomenology
of soft scattering.
It means, that the gBKFL dynamics predicts a substantial rise
with the energy and dipole size, $r$, of the diffraction slope, $B(\xi,r)$,
in accordance with
the energy and dipole size dependence of the effective
Regge slope, $\alpha_{eff}'(\xi,r)$ and due to a presence of the
geometrical component, $\propto r^{2}$, in (\ref{eq:17}) and
(\ref{eq:18}).

Generalized factorization formula (\ref{eq:12}) for the
forward diffraction slope can be re-written for somewhat
better understanding of anomalous properties
of the forward diffraction slope for production of
$V'(2S)$ vector mesons
%
%
\bea
B(\gamma^{*}\rightarrow V,\xi,Q^{2},{\bf q}=0) =
\frac{
\langle V |\sigma(\xi,r)B(\xi,r)|\gamma^{*}\rangle
}{
\langle V |\sigma(\xi,r)|\gamma^{*}\rangle} =
\nonumber \\
\frac{
\int\limits_{0}^{1} dz\int d^{2}{\bf{r}}\sigma(\xi,r)
B(\xi,r)\Psi_{V}^{*}({\bf{r}},z)\Psi_{\gamma^{*}}({\bf{r}},z)
}{
\int\limits_{0}^{1} dz\int d^{2}{\bf{r}}\sigma(\xi,r)
\Psi_{V}^{*}({\bf{r}},z)\Psi_{\gamma^{*}}({\bf{r}},z)}
\,
\label{eq:22}
\eea
%
%

%
%
\section{Diffraction slope for $\rho^{0}$
         and $\phi^{0}$ electroproduction:
         model predictions vs. experiment}

Firstly the model predictions for the diffraction slope
will be tested taking the fixed target and HERA data
of $V(1S)$ vector meson production.
The color dipole gBFKL dynamics predicts a substantial growth
with energy of the diffraction slope coming from Eqs.~(\ref{eq:18})
and (\ref{eq:21}).
According to simple geometrical behaviour, $\propto r^{2}$,
of the slope parameter (\ref{eq:18},\ref{eq:21}),
we expect a shrinkage of the diffraction slope with $Q^{2}$
in accordance with the scanning property in vector meson
production (see Eq.~(\ref{eq:2})).
In Fig.1 we compare the model predictions for 
$Q^{2}$ dependence of 
the diffraction
slope for $\rho^{0}$ production with the low energy
data of the CHIO \cite{CHIOrho} NMC \cite{NMCfirho}
and E665 \cite{E665rho}
collaborations and the data from
H1 \cite{H1rho96,H1rho96Q2,H1rho99Q2} and 
ZEUS \cite{ZEUSrho95,ZEUSrho97,ZEUSrho98,ZEUSrho95Q2,ZEUSrho99Q2} experiments.
Although the experimental data have still large error bars,
they show a trend to smaller values of the diffraction slope as $Q^{2}$
increases. 
We predict a steep shrinkage of $B(\rho^{0})$ with $Q^{2}$
on the scale $Q^{2}\in (0,5)$\,GeV$^{2}$:
it falls down, by $\sim 4~$\,GeV$^{-2}$ from $\sim 8.7$\,GeV$^{-2}$
at $Q^{2}=0$ down to 
5.0\,GeV$^{-2}$ at $Q^{2}=5$\,GeV$^{2}$ and
to 4.6\,GeV$^{-2}$ at $Q^{2}=10$\,GeV$^{2}$ 
in accordance with the low energy
CHIO, NMC and E665 data.
At HERA energy, we predict a higher shrinkage from $\sim
10.7$\,GeV$^{-2}$
at $Q^{2}=0$ down to $\sim 6.0$\,GeV$^{-2}$
at $Q^{2}=10$\,GeV$^{2}$ not in disagreement with the data of H1 and ZEUS
collaborations. 
Concerning the shrinkage of the diffraction slope with energy $W$, 
in the photoproduction limit $Q^{2}=0$ the data show a possible
presence of the considerably
large rise from the fixed target to HERA energy range.
However, the large error bars of the data affect the large
errors on $\alpha'$- fit \cite{ZEUSrho98} and preclude any definitive
statement. 
In Fig.2 we predict this substantial growth, 
by $\sim 2.3-2.4$\,
GeV$^{-2}$ from $\sim 8.3-8.4$\,GeV$^{-2}$
at $W=10$\,GeV up to $\sim 10.7$\,GeV$^{-2}$
at $W=100$\,GeV in accordance with the data from the fixed target
experiments \cite{FTrho} and the data from HERA experiments
\cite{H1rho96,ZEUSrho95,ZEUSrho97,ZEUSrho98}. This rise corresponds to
effective Regge slope, $\alpha'\sim 0.25-0.26$\,GeV$^{-2}$.

We would like to emphasize, that the overall effective
Regge slope, $\alpha'$, contains the energy dependent contribution
of the perturbative component, $\alpha_{eff}'(xi,r)$, characterizing
the energy rise of the gBFKL slope, $B_{pt}(\xi,r)$
(see Eq.~(\ref{eq:17},(\ref{eq:21})), and the constant nonperturbative
(soft) Regge slope, $\alpha_{npt}'= 0.15$\,GeV$^{-2}$, corresponding
to the soft component of the slope, $B_{npt}(\xi,r)$
(see Eq.~(\ref{eq:18})).
As was mentioned in Ref.~\cite{NZZslope}, 
in the energy range, $W\in
(50-200)$\,GeV, the effective
Regge slope, $\alpha_{eff}'(\xi,r)$, varies slowly within
the interval $\sim (0.15-0.20)$\,GeV$^{-2}$ at different
scanning radii $\lsim 1$\,fm and
is approximately a flat function of
the scanning radius at fixed energy 
corresponding to HERA experiments;
for instance, at $W=100$\,GeV, 
$\alpha_{eff}'\sim 0.15$\,GeV$^{-2}$ at $r_{S}\sim 0.1$\,fm\,,
$\alpha_{eff}'\sim 0.16-0.17$\,GeV$^{-2}$ at $r_{S}\sim 0.2-0.5$\,fm\,,
$\alpha_{eff}'\sim 0.19.-0.20$\,GeV$^{-2}$ at $r_{S}\sim 0.6-0.9$\,fm\,,
$\alpha_{eff}'\gsim 0.20$\,GeV$^{-2}$ at $r_{S}\gsim 1.0$\,fm\,.
$\alpha_{npt}'$ only slightly modifies the overall effective
Regge slope $\alpha'$.

In Fig.2 we show also the energy dependence of the slope parameter
for $\rho^{0}$ virtual photoproduction at $Q^{2}\sim 10$\,GeV$^{2}$
vs. NMC \cite{NMCfirho} and H1 \cite{H1rho96Q2,H1rho99Q2} data.
The growth with energy $W$ is much smaller than at $Q^{2}=0$:
$B(\rho^{0})$ rises from $\sim 4.4$\,GeV$^{-2}$ at $W=10$\,GeV up to
$\sim 6.0$\,GeV$^{-2}$ at $W=100$\,GeV.
It correspond to effective Regge slope $\sim 0.17$ GeV$^{-2}$.
At $Q^{2}\sim 20-30$ GeV$^{2}$, we predict $\alpha'\sim 0.15$
GeV$^{-2}$, which is
in accordance with value of the effective shrinkage rate
of the diffraction slope for $J/\Psi$ elastic photoproduction
($Q^2=0$) presented in the paper \cite{NNPZZ98}.
It confirms an approximate flavor independence of the effective
Regge slope in the scaling variable $Q^{2}+m_{V}^{2}$.

Fig.3 shows the analogical $W$ dependence of the slope for real 
$\phi^{0}$ photoproduction together with the data from fixed
target \cite{Philownu,Philownuold} and collider
HERA experiments \cite{ZEUSphi96}.
Unfortunately, the error bars are quite a large to
see a clear evidence of the shrinkage of $B(\phi^{0})$
with energy. 
The model predictions do not show a deviation from the data
and it is not in disagreement with the conclusion 
about a shrinkage of the diffraction peak with energy
expected from the gBFKL dynamics.
The energy growth of $B(\phi^{0})$ on the interval
of $W\in (10-100)$\,GeV, is expected to correspond 
to overall effective
Regge slope $\alpha'\sim 0.20-0.21$\,GeV$^{-2}$.

Regarding a comparison with the data, the most
straightforward theoretical predictions are
for the forward production  and we calculate
$d\sigma/dt|_{t=0}$ and $B(t=0)$. 
The data on the vector meson production correspond to a slope
extracted over quite a broad range of $t$ using an
extrapolation to $t=0$,
and the minimal value of $t$, corresponding to the fist experimental
point in $t$- distribution, is relatively far from $t=0$.
Also the range of $t$ is different in different experiments.
This fact explains
quite a large dispersion of the low-energy data which is
the most striking for $\phi^{0}$ production depicted on Fig.3
(see also Fig.2).
Moreover, the above extrapolation
is not always possible and one often reports the
$t$-integrated production cross sections. 
Because of the model calculations
are at $t=0$ and because of a well known rapid rise of the
diffraction slope towards $t=0$ \cite{Schiz},
the experimental data may
underestimate $B(V)$ at $t=0$.
For average $\langle t \rangle \sim$
0.1-0.2\,GeV$^{2}$ which dominate the integrated total
cross section, the diffraction slope is smaller than
at $t=0$ by $\sim 1$\,GeV$^{-2}$ \cite{Schiz}. We take these $\pi N$
scattering data for the guidance, and for more direct comparison with
the presently available experimental data instead of the directly
calculated $B(t=0)$ we report  in Figs.1-3 the value
%
%
\beq
B=B(t=0)-1\,GeV^{-2}
\label{eq:23}
\eeq
%
%
The uncertainties in the value of $B$ and with this
evaluation
(\ref{eq:23})
presumably do not exceed
$10\%$ and can be reduced when more accurate data will
become available.
However, hereafter we will present the model predictions for the
diffraction slope at $t=0$. 

More detailed predictions for the energy and $Q^{2}$ dependence
of the forward diffraction slope $B(V,t=0)$ for the $\rho^{0}$
and $\phi^{0}$ production (for $T$, $L$ and mixed
$T + \epsilon L$ polarizations, with $\epsilon=1$)
are presented in Fig.4.
They show a substantial shrinkage of the
elastic peak with energy at different $Q^{2}$.
The energy rise of the diffraction slope
is more evident than for the production
of heavy vector mesons \cite{NNPZZ98}.

The rate of rise with energy of the diffraction slope
decreases slowly with $Q^{2}$:
on the interval of the c.m.s. energy $W\in (10-100)$\,GeV
the corresponding $\alpha'\sim 0.25$\,GeV$^{-2}$
at $Q^{2}=0$, $\alpha'\sim 0.21$\,GeV$^{-2}$
at $Q^{2}\sim 0.5$\,GeV$^{2}$,
$\alpha'\sim 0.19$\,GeV$^{-2}$ at $Q^{2}\sim 1.0$\,GeV$^{2}$,
$\alpha'\sim 0.17$\,GeV$^{-2}$ at $Q^{2}\sim 5.0$\,GeV$^{2}$ and
$\alpha'\sim 0.16$\,GeV$^{-2}$ at $Q^{2}\sim 20$\,GeV$^{2}$.
The effective Regge slope
becomes still smaller at very large $Q^{2}$ and $W$
when the scanning
radius $r_{S}\lsim R_{c}$ and
a contribution of $\alpha_{npt}'=0.15$\,GeV$^{-2}$ 
to overall $\alpha'$ becomes 
practically insignificant. 
At $Q^{2}\gsim 100$\,GeV$^{2}$ 
when the scanning radius $r_{S}\lsim R_{c}$,
and one can observe a standard picture of a decreasing rate of
energy growth of $B(V)$ expected from gBFKL dynamics
(see Fig.~4).

The above results for the energy growth of the slope parameter
can be tested in higher statistics data from HERA experiments
measuring the exclusive electroproduction of vector
mesons. The measurement of energy rise of the slope
parameter at different $Q^{2}$ can give an information
about a contribution of the nonperturbative component
of the diffraction slope, $B_{npt}(\xi,r)$, and the
effective Regge slope, $\alpha_{eff}'(\xi,r)$.
The more precise data
could also test the universal properties
of diffraction slope and effective Regge slope for production
of different vector mesons, i.e. a similarity between
the production of different vector mesons when compared
at the same value of the scanning radius $r_{S}$ and/or the
same value of $Q^{2}+m_{V}^{2}$ (see Eq.~(\ref{eq:2})).
Such a comparison must be performed at the same energy
and/or rapidity $\xi$, which also means the equality of
$x_{eff}$ at equal $Q^{2}+m_{V}^{2}$
(see \cite{NNPZ97,NZZslope}).
The value of $Q^{2}$ must be large enough so that the scanning
radius $r_{S}$ is smaller than the radii of vector mesons, 
$r_{S}\lsim R_{V}$.
It means, that for all reactions $\gamma^{*}~p\rightarrow V~p$
with the same $r_{S}$ and $\xi$, we predict approximately the same
$B(V)$ and $\alpha_{eff}'$ \cite{NNPZZ98}.

Although a new data on the diffraction slope were obtained
from collider HERA experiments 
measuring the real (virtual) photoproduction of 
vector mesons,
the present experimental information on the energy and $Q^{2}$
dependence of the diffraction slope 
for vector meson production is not still very conclusive.
Especially, it concerns to
$J/\Psi$ photoproduction.
There are no data yet on the diffraction slope
for the real (virtual) photoproduction of $\Upsilon$ and
the radially excited $(2S)$ heavy vector
mesons\footnote{
More detailed discussion of the data on the slope
parameter for heavy vector meson production is presented
in Ref.~\cite{NNPZZ98}}.
The data on the diffraction slope measuring the photo-
and electroproduction of light vector mesons presented on
Figs.~1-3 have still large error bars.
The ZEUS and H1 data on virtual photoproduction give
$B(\rho^{0},W\sim 80$\,GeV,$7<Q^{2}<25$\,GeV$^{2})=
5.1+1.2-0.9(stat)\pm 1.0(syst)$\,GeV$^{-2}$ \cite{ZEUSrho95Q2},
$B(\rho^{0},W\sim 100$\,GeV,$Q^{2}=28$\,GeV$^{2})=
4.4+3.5-2.8(stat)+3.7-1.2(syst)$\,GeV$^{-2}$ \cite{ZEUSrho99Q2}
and
$B(\rho^{0},W\sim 75$\,GeV,$Q^{2}=21.2$\,GeV$^{2})=
4.7\pm 1.0(stat)\pm 0.7(syst)$\,GeV$^{-2}$ \cite{H1rho99Q2} 
which is close to 
$B(J/\Psi,W=90$\,GeV$,Q^{2}=0)=
4.7\pm 1.9$\,GeV$^{-2}$,
$B(J/\Psi,W=90$\,GeV$,Q^{2}=0)=
4.0\pm 0.3$\,GeV$^{-2}$ from H1 data \cite{H1Psi94,H1Psi96}
and to
$B(J/\Psi,W=90$\,GeV$,Q^{2}=0)=
4.5\pm 1.4$\,GeV$^{-2}$ 
$B(J/\Psi,W=90$\,GeV$,Q^{2}=0)=
4.6\pm 0.4(stat)+0.4-0.6(syst)$\,GeV$^{-2}$ 
from ZEUS data \cite{ZEUSPsi95,ZEUSPsi97}
in accordance with $(Q^{2}+m_{V}^{2})$- scaling of the
diffraction slope.
High statistics data are needed from the both fixed target
and the collider HERA experiments for both
the exploratory study of very interesting $Q^{2}$ and energy
dependence of $B(V)$ and 
the precise test of the $(Q^{2}+m_{V}^{2})$- 
scaling of the diffraction slope.

%
%
\section{Anomalous diffraction slope
         in electroproduction of $2S$ 
         radially excited vector mesons}

Now we concentrate on the production of radially excited
$V(2S)$ light vector mesons, where the node effect is
known to be presented 
- the $Q^{2}$ and energy dependent cancellations
from the soft (large size) and hard (small size)
contributions, i.e. from the region
above and below the node position, $r_{n}$,
to the $V'(2S)$ production amplitude.
The strong $Q^{2}$ dependence of these cancellations comes
from the scanning phenomenon (\ref{eq:2}) when
the scanning radius $r_{S}$ for some value of $Q^{2}$
is close to $r_{n}\sim R_{V}$.
The energy dependence of the node effect comes from the 
different energy dependence of the dipole cross section
at small ($r<R_{V}$) and large ($r>R_{V}$) dipole sizes.
The strong node effect in production of radially
excited light vector mesons leading to an anomalous
$Q^{2}$ and energy dependence of the production
cross section has been demonstrated in Ref.~\cite{NNPZ97}\footnote{
Manifestations of the node
effect in electroproduction on nuclei were discussed earlier, see
\cite{NNZanom} and \cite{BZNFphi}}

Note, that the predictive power is weak
and is strongly model dependent in the region of $Q^{2}$ and energy
where the node effect becomes exact.

For the production of $V'(2S)$ light vector mesons, 
the node effect depends on the
polarization of the virtual photon and of the produced vector
meson \cite{NNPZ97}. 
The wave functions of $T$ and $L$ polarized (virtual)
photon are different.
Different regions of $z$ contribute to the
${\cal M}_{T}$ and ${\cal M}_{L}$.
Different scanning radii
for production of $T$ and $L$ polarized vector mesons
and different energy dependence of $\sigma(\xi,r)$ at
these scanning radii
lead to a different $Q^{2}$ and energy dependence of the
node effect in production of $T$ and $L$ polarized 
$V'(2S)$ vector mesons.
Not so for production of heavy quarkonia,
where the node effect is very weak and
is approximately polarization independent.
However, there is a weak polarization dependence of the
node effect for $\Psi'$ photoproduction \cite{NNPZZ98} and
this weak node effect still leads to a nonmonotonic $Q^{2}$
dependence of the diffraction slope.
For $\Upsilon'$ production the node effect is negligibly small
and is polarization independent with very high accuracy.

There are two possible scenarios for the node effect:
the undercompensation
and the overcompensation regime \cite{NNZanom}.
In the undercompensation case,
the $2S$ production amplitude
$\langle V'(2S)|\sigma(\xi,r)|\gamma^*\rangle$
is dominated by the positive contribution coming from small
dipole sizes, $r\lsim r_{n}$ ($r_{n}$ is the node position),
and the $V(1S)$ and $V'(2S)$ photoproduction
amplitudes have the same sign. 
This scenario corresponds namely to the production
of $2S$ heavy vector mesons, $\Psi'(2S)$ and $\Upsilon'(2S)$.
In the overcompensation case,  
the $2S$ production amplitude
$\langle V'(2S)|\sigma(\xi,r)|\gamma^*\rangle$
is dominated by the negative contribution coming from large
dipole sizes, $r\gsim r_{n}$, 
and the $V(1S)$ and $V'(2S)$ photoproduction
amplitudes have the opposite sign. 

The anomalous properties of the diffraction slope can
be understood
from the expression (\ref{eq:22}).
The denominator represents the well known production
amplitude $\langle V(V')|\sigma(\xi,r)|\gamma^*\rangle$.
As it was mentioned, 
the $1S$ production amplitude
is dominated by contribution from
dipole size $r\sim r_{S}$ (\ref{eq:2}).
However, 
due to $\propto r^{2}$
behaviour of the slope parameter (see (\ref{eq:18}) and
(\ref{eq:21})),
the integrand of the matrix element in the numerator,
$\langle V(1S)|\sigma(\xi,r)B(\xi,r)|\gamma^*\rangle$,
is $\sim r^{5}\exp(-\epsilon r)$ and is peaked by
$r\sim r_{B} = 5/3r_{S}$.

Let us start from $T$ polarized $\rho'(2S)$ 
In Ref.~\cite{NNPZ97}
using our model wave functions for vector mesons, we found 
the undercompensation scenario at $Q^{2}=0$
for the production amplitude
$\langle V_{T}'(2S)|\sigma(\xi,r)|\gamma^*\rangle$, 
which is positive valued.
However, because of the large numerical factor $\sim 10$ for 
$r_{B}\sim 10/\sqrt{Q^{2}+m_{V}^{2}} > r_{S}$,
the matrix element
$\langle V_{T}'(2S)|\sigma(\xi,r)B(\xi,r)|\gamma^*\rangle$ 
in the numerator of Eq.~(\ref{eq:22})
corresponds to the overcompensation scenario
at $Q^{2}=0$ and at energy range
$\lsim 15-20\,$ GeV and is negative valued.
As the result,
the numerator and denominator have the opposite signs
resulting in a negative value for the diffraction slope,
$B(V_{T}'(2S))$ at $Q^{2}=0$.
However the node effect for production of $phi'(2S)$ is
weaker resulting in positive valued numerator of Eq.~(\ref{eq:22}).
Both the numerator and denominator have the same sign and
we start with positive valued diffraction slope.
Such a situation is depicted in Fig.~5 (bottom boxes) for both the
$\rho'(2S)$ and $\phi'(2S)$ production,
where we present the model predictions
for the forward diffraction slope $(t=0)$ as a function of $Q^{2}$
at different values of the c.m.s. energy $W$.

A decrease of the scanning radius with $Q^{2}$ leads to a very
rapid decrease of the negative contribution to the
diffraction slope
coming from $r\gsim r_{n}$ and consequently, 
leads to a steep rise of the
negative valued $B(V_{T}'(2S))$
with $Q^{2}$ for $\rho'(2S)$ production
(positive valued $B(V_{T}'(2S))$
with $Q^{2}$ for $\phi'(2S)$ production).
For $\rho'(2S)$ production 
at some value of $Q^{2}\sim Q^{2}_{T}\sim 0.01$\,GeV$^{2}$,
one encounters the exact cancellation of the large and small
distance contributions, i.e. the exact node effect for
the numerator of Eq. (\ref{eq:22}), and 
$B(V_{T}'(2S)) = 0$. Not so for $\phi'(2S)$ production,
where the numerator of Eq.~(\ref{eq:22}) is in the 
undercompensation regime already at very small energies
$\sim 5\,$ GeV. 

We would like to emphasize that the position of $Q^{2}_{T}$
is model dependent and can be shifted towards to smaller
or to larger values.

At larger $Q^{2}$ and smaller scanning
radius, one enters the undercompensation scenario also for
numerator
$\langle V_{T}'(2S)|\sigma(\xi,r)B(\xi,r)|\gamma^*\rangle$.
Thus, the diffraction slope will be positive valued and continues to
rise strongly with $Q^{2}$ due to a more rapid decrease with $Q^{2}$
of the negative contribution to the slope parameter coming
from $r\gsim r_{n}$ in numerator than in denominator
(the numerator has much stronger node effect than the
denominator).
At still larger $Q^{2}$, i.e. smaller
scanning radii $r_{S}$, the node effect also for the
numerator becomes to be weaker and as the result,  
the slope parameter 
at fixed energy $W$ and some value of $Q^{2}\sim (0.5-2.0)$\,GeV$^{2}$,
has a maximum of $B(V_{T}'(2S))$.
At very large $Q^{2}\gg m_{V}^{2}$, when the node effect becomes
negligible, $B(V_{T}'(2S))$ has the standard $Q^{2}$- behaviour
and decreases monotonously with $Q^{2}$
following the behaviour of the diffraction slope $B(V_{L,T}(1S))$
for $V(1S)$ mesons (see Fig.~4).

The more interesting situation is for production of $L$ polarized
$V'(2S)$ mesons resulting in a very spectacular pattern of $Q^{2}$
dependence of the slope parameter shown in Fig.~5 (middle boxes).
Using our model wave functions, we predicted overcompensation
for the production amplitude  
$\langle V_{L}'(2S)|\sigma(\xi,r)|\gamma^*\rangle$ 
\cite{NNPZ97}.
Because of $r_{B}> r_{S}$, the matrix element
$\langle V_{L}'(2S)|\sigma(\xi,r)B(\xi,r)|\gamma^*\rangle$,
will be also in the overcompensation regime. 
For this reason, the both matrix elements have the same sign
and according to (\ref{eq:22}) the slope parameter
$B(V_{L}'(2S))$ will be positive valued at
$Q^{2}=0$ (see Fig.~5).
Consequently, with the
decrease of the scanning radius with $Q^{2}$,
there is a
rapid decrease of the negative contributions 
to the numerator and denominator
of Eq.~(\ref{eq:22}) coming from $r\gsim r_{n}$.
For some $Q^{2}\sim Q'^{2}_{L}\sim 0.5-1.0$\,GeV$^{2}$
one encounters the exact node effect firstly for the denominator
due to $r_{B}> r_{S}$. This fact corresponds to
a presence of the peak for $B(V_{L}'(2S))$ for both the $\rho_{L}'(2S)$ and
$\phi_{L}'(2S)$ production. The value of $B(V_{L}'(2S))$ corresponding to
this exact node effect will be finite due to a different node effect
for the real and imaginary part of the production amplitude.
This fact also reflects the
continuous transition of $B(V_{L}'(2S))$ from positive to 
negative values
when the matrix element in denominator passes from the
overcompensation to undercompensation regime.
Thus, for $Q^{2}\geq Q'^{2}_{L}\sim 0.5-1.0$\,GeV$^{2}$, 
the denominator will be in
the undercompensation regime and $B(V_{L}'(2S))$ starts to
rise from its minimal negative value.
Note, that the numerator is still in the overcompensation.
The further pattern of the $Q^{2}$ behaviour of 
$B(V_{L}'(2S))$ is analogical to that for
$Q^{2}$ dependence of $B(V_{T}'(2S))$.
However,  the exact node effect for
the numerator in Eq.~(\ref{eq:22}), resulting in $B(V_{L}'(2S))=0$,
will be at $Q^{2}\sim Q^{2}_{L} > Q'^{2}_{L}$.
For $\phi'(2S)$ production because of different node effect for
the real and imaginary part of production amplitude,
at HERA energy range $W\sim 50-200\,$ GeV 
$B(V_{L}'(2S))$ never reaches the zero value corresponding
to the exact node effect for the numerator of Eq.~(\ref{eq:22}).

For the production of polarization unseparated $V'(2S)$,
the anomalous properties of $B(V_{L}'(2S))$ are essentially
invisible and the corresponding slope parameter
$B(V'(2S))$ is shown in Fig.~5 (bottom boxes).
Although the above
value of $Q_{T}^{2}$ is too small to be measured experimentally
(we can not exclude that $Q^{2}$ dependence of $B(V_{T}'(2S))$
will start from positive valued $B(V_{T}'(2S))$ at small energies
also for $\rho'(2S$) production),
we predict nonmonotonic $Q^{2}$ dependence of the diffraction slope
for production of $T$ polarized 
and polarization unseparated $\rho'(2S)$ and $\phi'(2S)$,
strikingly different from monotonic $Q^{2}$ behaviour
of the slope parameter for $V(1S)$ production (see
Fig.~4).
For production of $\rho_{L}'(2S)$ and $\phi_{L}'(2S)$,
we predict anomalous $Q^{2}$ behaviour of $B(V_{L}'(2S))$.
Here we can not insist on the precise values of $Q^{2}_{T}$, 
$Q^{2}_{L}$ and $Q'^{2}_{L}$ which
is subject of the soft-hard cancellations.
We would like to only emphasize that the exact node effect
for $B(V_{T}'(2S))$ and $B(V_{L}'(2S))$
is at a finite $Q^{2}_{T}$ and $Q^{2}_{L}$, $Q'^{2}_{L}$ 
respectively.
Such a nonmonotonic $Q^{2}$ dependence of $B(V_{T}'(2S))$ and/or
$B(V'(2S))$ can
be tested experimentally at HERA measuring the virtual
photoproduction of the $\rho'(2S)$ and $\phi'(2S)$ at
$Q^{2}\in (0-10)$\,GeV$^{2}$.
The above discussed anomalous $Q^{2}$ dependence of $B(V_{L}'(2S)$
could be also investigated at HERA separating $L$ polarized
$\rho_{L}'(2S)$ and $\phi_{L}'(2S)$ at moderate 
$Q^{2}\sim (0.1-2.0)$\,GeV$^{2}$. 
Here we would like to emphasize that only the experiment can help
in decision between the undercompensation and overcompensation
scenarios which affect the anomalous properties of the production
cross section and diffraction slope.

The energy dependence of the slope parameter $B(V'(2S))$ 
at different $Q^{2}$ is shown
in Fig.6 and has its own peculiarities.
Let us start with $B(V_{T}'(2S))$ at $Q^{2}=0$.
Fig.~6 demonstrates (top boxes) steeper rise with energy 
of the diffraction slope at lower $Q^{2}$.
There are several reasons for such a behaviour.
First,
the gBFKL dynamics predicts a steeper rise with energy of
the positive contribution to the $2S$ amplitudes
$\langle V'(2S)|\sigma(\xi,r)|\gamma^*\rangle$ and
$\langle V'(2S)|\sigma(\xi,r)B(\xi,r)|\gamma^*\rangle$
coming from small size dipoles $r\lsim r_{n}$ 
than the negative contribution coming from 
large size dipoles $r\gsim r_{n}$.
Thus, 
the destructive interference of these two contributions
is weaker at higher energy.
Second,
at $Q^{2}=0$,
the denominator of Eq.~(\ref{eq:22}) is in the undercompensation,
whereas the numerator is in the overcompensation regime
(numerator is in the undercompensation regime for 
$\phi'(2S)$ production) 
and the corresponding scanning radii for the numerator
and denominator are different, $r_{B}> r_{S}$.
Third, the energy dependence of the slope parameter
is given by the effective Regge slope $\alpha'$.
Thus, the above destructive interference 
in numerator 
decreases drastically with $W$
the negative contribution from $r\gsim r_{n}$ 
until the exact node effect is reached, i.e.
$B(V_{T}'(2S))=0$, and the undercompensation scenario
also for the numerator of Eq.~(\ref{eq:22})
starts to be realized at $W\sim 20$\,GeV.
However, closeness of the node position in
the numerator of Eq.~(\ref{eq:22}) leads to
a small negative value of $B(V_{T}'(2S))$
at $W\sim 5\,$ GeV and as a result
it leads in a little bit steeper growth with energy
of $B(V_{T}'(2S))$ than the expected 
energy rise of the slope coming only 
from the effective Regge slope. 
For example, for $\rho'(2S)$ production 
we predict the rise of $B(V_{T}'(2S))$, by $\sim 2.8$\,GeV$^{-2}$,
from $W=10$ to $100$\,GeV. 
At $Q^{2}\gsim 1.0$\,GeV$^{2}$, 
when both the numerator and denominator are in the
undercompensation regime and the node effect becomes weak,
the energy growth of $B(V'(2S))$ is
connected mainly with the effective Regge slope and we
predict approximately the same quantities and energy growth
for $B(V'(2S))$ and
$B(V(1S))$ (compare Fig.~4 and Fig.~6).

The successful separation of the longitudinally polarized
$V_{L}'(2S)$ mesons at HERA offers an unique possibility
to study an anomalous $Q^{2}$ and energy dependence of
the diffraction slope connected with the overcompensation scenario
of the denominator of Eq.~(\ref{eq:22}).
At $Q^{2}=0$, we have onset of the overcompensation scenario
for both the numerator and denominator of Eq.~(\ref{eq:22}).
At moderate energy and $Q^{2}$ closed but smaller than
$Q'^{2}_{L}\sim 0.5$\,GeV$^{2}$, the negative contribution
coming from $r\gsim r_{n}$ still takes over in the denominator
(the numerator is safely in the overcompensation
regime due to $r_{B}> r_{S}$). 
Due to a steeper rise with energy of
the positive contribution to the $2S$ production amplitude
coming from small size dipoles $r\lsim r_{n}$ 
than the negative contribution coming from 
large size dipoles $r\gsim r_{n}$, we find an exact cancallation
of these two contributions to the denominator and a maximum
of the diffraction slope $B(V_{L}'(2S))$ at some intermediate energy
followed by a rapid continuous transition
from the positive to negative values,
when the matrix element in denominator of Eq.~(\ref{eq:22}) passes
from the overcompensation to the undercompensation regime.
Different node effect for the real and imaginary part of 
the production amplitude provides such a continuous transition.
Then, at larger energies, 
the production amplitude is in the undercompensation regime,
$B(V_{L}'(2S))$ is negative valued and starts to rise 
from the minimal negative value. 
This situation is depicted in Fig.~6 (middle boxes), where we
predict with our model wave functions 
such a nonmonotonic energy behaviour
of $B(V_{L}'(2S))$ for both $\rho'(2S)$
and $\phi'(2S)$ production at $Q^{2}\lsim 0.7-1.0$\,GeV$^{2}$.
The position $W_{t}$ 
of maximum and the transition from the positive to negative 
values of the logitudinally polarized diffraction slope depends
on $Q^{2}$. For example, at $Q^{2}\sim 0.7$\,GeV$^{2}$,
we find $W_{t}\sim 70-80$\,GeV. Then, the position of $W_{t}$
is shifted to smaller values of $W$ at larger
$Q^{2}\gsim 0.7$\,GeV$^{2}$ at can be measured at HERA.

At higher $Q^{2}$ and smaller scanning radii,
the further pattern of the energy behaviour of 
$B(V_{L}'(2S)$ is an analogical
to $W$ dependence of $B(V_{T}'(2S))$.
At still larger $Q^{2}$, after
the exact node effect was reached also in the numerator
of Eq.~(\ref{eq:22}) at $\sim Q^{2}_{L}>Q'^{2}_{L}$,
both the numerator and denominator are in the
undercompensation regime. 
Consequently, the node effect also in the
numerator starts to be weaker with $Q^{2}$ and 
the energy growth of $B(V'_{L}(2S))$ is
controlled practically 
by the effective Regge slope.
As the result we
predict again almost the same quantities and energy growth
for $B(V'_{L}(2S))$ and $B(V_{L}(1S))$.

If the leptoproduction of the transversally and longitudinally
polarized $V_{T}'(2S)$ and $V_{L}'(2S)$ mesons will be
separated experimentally, there is a
possibility of experimental
determination of a concrete scenario in $T$ and $L$
polarized $2S$ production
amplitude by a measurement of the corresponding diffraction
slopes at $t=0$ and at $Q^{2}=0$, where the node
effect is found to be the strongest.
If at the same energy and $Q^{2}=0$
the slope parameter for $V_{T}'(2S)$ production will be
smaller (it can be also negative valued)
than the corresponding slope parameter for
$V_{T}(1S)$ production, then the $2S$ production amplitude
is in the undercompensation regime.
In the opposite case, if $B(V_{T}'(2S))> B(V_{T}(1S)$,
then the corresponding $T$ polarized $2S$ amplitude is
in the overcompensation. 
The analogical conclusion concerns to $L$ polarized
$2S$ production amplitude, where the values of $Q^{2}$
should be high enough to have the data with a reasonable
statistics, however must not be very large in order
to have a strong node effect. We propose the range
of $Q^{2}\in 0.5-2.0$\,GeV$^{2}$ for a possible study
of the overcompensation scenario at HERA.
The further supplementary
indication of the overcompensation scenario is
assumed to be an existence of the maximum
and/or minimum of the diffraction slope and
subsequent a sudden rise of $B(V'(2S))$ at some nonzero value
of $Q^{2}$.




\section{Conclusions} 

We study the 
diffractive photo- and electroproduction
of ground state $1S$ and radially excited $2S$ 
vector mesons within the color dipole gBFKL dynamics
with the main emphasis related to the  
diffraction slope.
There are two main consequences
of vector meson production
coming from the gBFKL dynamics.
First, the energy dependence of the $1S$ vector meson production
is controlled by the energy dependence of the dipole cross
section which is steeper for smaller dipole sizes.
The energy dependence of the diffraction slope for $V(1S)$
production is given by the effective Regge slope with a small
variation with energy.
Second
the $Q^{2}$ dependence of the $1S$ vector meson production is
controlled by the shrinkage of the transverse size of the virtual
photon and the small dipole size dependence of the color dipole cross
section.
The $Q^{2}$ behaviour of the diffraction slope is given by
the simple geometrical properties, $\sim r^{2}$, coming
from the gBFKL phenomenology
of the slope parameter.
In the gBFKL dynamics, we expect a fast subasymptotic shrinkage
of the diffraction cone from the CERN/FNAL to HERA energy 
due to intrusion of large distance effects. 
We have predicted a reach pattern of 
$Q^{2}$ and energy dependence
of the diffraction slope for the $\rho^{0}$ and $\phi^{0}$ production
and find a substantial rise (by $\sim 2.3-2.4$\,GeV$^{-2}$ 
for $\rho^{0}$ production
and by $\sim 1.9-2.0$\,GeV$^{-2}$ for $\phi^{0}$ production)
from the fixed target, $W\sim 10-15$\,GeV, to
the collider HERA, $W\sim 100-150$\,GeV, range of energy.
The model predictions for the diffraction slope 
for the $\rho^{0}$ and $\phi^{0}$ production are in 
agreement with 
the data from the fixed target 
(CHIO, NMC) and collider HERA (H1, ZEUS) experiments.
However, the relatively large error bars of the data preclude
any definite statement about a shrinkage of the slope
parameter with energy. The data show a trend to smaller
values of the diffraction slope as $Q^{2}$ increases.

The second class of predictions is related to the
diffraction slope for the production of $2S$
vector mesons.
As a 
consequence of the strong node effect
in electroproduction of $2S$ light vector
mesons $\rho'(2S)$ and $\phi'(2S)$, 
we present the strong case for the anomalous $Q^{2}$ and energy
dependence of the diffraction slope at $t=0$.
We find a nonmonotonic $Q^{2}$ dependence of the slope
parameter which can be tested at HERA in the range
of $Q^{2}\in (0-10)$\,GeV$^{2}$ measuring
the virtual photoproduction of $\rho'(2S)$ and $\phi'(2S)$
mesons.
For the production of longitudinally polarized $2S$
mesons, the production amplitude is in the
overcompensation scenario and
we find a very rapid transition of the slope parameter
$B(V_{L}'(2S))$ from positive to negative values
at $Q^{2}= Q'^{2}_{L}\sim 0.5-2.0$\,GeV$^{2}$ 
as a consequence of a reaching of the exact
node effect by passing from the overcompensation
to undercompensation scenario in $2S$ production amplitude.
The position of this rapid transition, $Q'^{2}_{L}$, 
is energy dependent and leads
to nonmonotonic energy dependence of $B(V_{L}'(2S))$ at fixed
$Q^{2}$. 

At $Q^{2}=0$, when 
the node effect is strong, for undercompensation
scenario we predict
smaller $B(V_{T}'(2S))$ than $B(V_{T}(1S))$.
However, for overcompensation
scenario we predict larger $B(V_{L}'(2S))$ than $B(V_{L}(1S))$.
This is a very crucial point of a possible experimental
determination of a concrete scenario measuring 
(and the position of the node as well) the diffraction
slope at $t=0$
for the production of $V'(2S)$ mesons in the photoproduction limit. 

At larger $Q^{2}$ and/or shorter scanning radius, the node effect
becomes weak and we predict for $V'(2S)$
mesons the standard monotonic $Q^{2}$
and energy dependence of the slope parameter like for $V(1S)$
mesons.
One needs the higher accuracy data from the both fixed target
and the collider HERA experiments for
the exploratory study of $Q^{2}$ and energy
dependence of the diffraction slope at $t=0$.  


\pagebreak
{\bf Figure captions:}
\begin{itemize}

\item[Fig.~1]
~- The color dipole model
predictions for the $Q^{2}$ dependence of the
diffraction slope for the production of $\rho^{0}$
vs. the low-energy fixed target CHIO \cite{CHIOrho},
NMC \cite{NMCfirho}, E665 \cite{E665rho}
and high-energy  ZEUS
\cite{ZEUSrho95,ZEUSrho97,ZEUSrho98,ZEUSrho95Q2,ZEUSrho99Q2} and H1 
\cite{H1rho96,H1rho96Q2,H1rho99Q2} data.

\item[Fig.~2]
~- The color dipole model
predictions for the $W$ dependence of the
diffraction slope for the production of $\rho^{0}$
vs. the low-energy fixed target \cite{FTrho,NMCfirho},
and high-energy  ZEUS
\cite{ZEUSrho95,ZEUSrho97,ZEUSrho98} and H1 
\cite{H1rho96,H1rho96Q2,H1rho99Q2} data
The top solid curve is a prediction for the diffraction
slope at $Q^{2}=0$.
The lower dashed curve represents a prediction at $Q^{2}=10$\,GeV$^{2}$.   

\item[Fig.~3]
~- The color dipole model
predictions for the $W$ dependence of the diffraction slope
for the real photoproduction of
$\phi^{0}$
vs. the low-energy fixed target \cite{Philownu,Philownuold}
and high-energy  ZEUS data
\cite{ZEUSphi96}.

\item[Fig.~4]
~- The color dipole model
predictions for the $W$ dependence of the diffraction slope $B(t=0)$
for production of
transversely (T)
(top boxes), longitudinally (L)
(middle boxes) polarized 
and polarization-unseparated
(T) + $\epsilon$(L) (bottom boxes)
$\rho^{0}$ and $\phi^{0}$
for $\epsilon = 1$ 
at different values of $Q^2$. 

\item[Fig.~5]
~- The color dipole model
predictions for the $Q^{2}$ dependence of the diffraction slope $B(t=0)$
for production of
transversely (T)
(top boxes), longitudinally (L)
(middle boxes) polarized 
and polarization-unseparated
(T) + $\epsilon$(L) (bottom boxes)
$\rho'(2S)$ and $\phi'(2S)$
for $\epsilon = 1$ 
at different values of the c.m.s. energy $W$. 

\item[Fig.~6]
~- The color dipole model
predictions for the $W$ dependence of the diffraction slope $B(t=0)$
for production of
transversely (T)
(top boxes), longitudinally (L)
(middle boxes) polarized 
and polarization-unseparated
(T) + $\epsilon$(L) (bottom boxes)
$\rho'(2S)$ and $\phi'(2S)$
for $\epsilon = 1$ 
at different values of $Q^2$. 

\end{itemize}

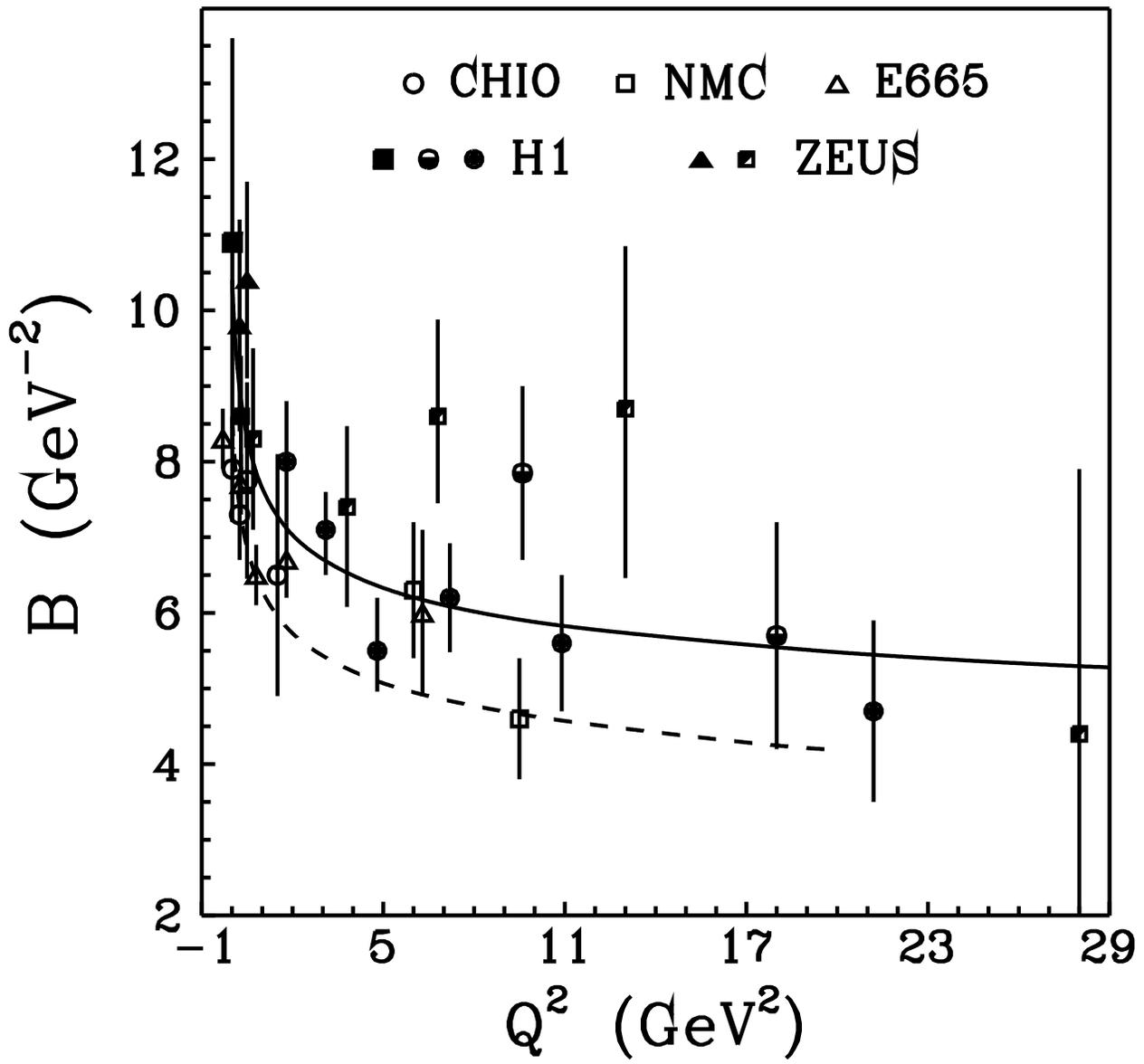
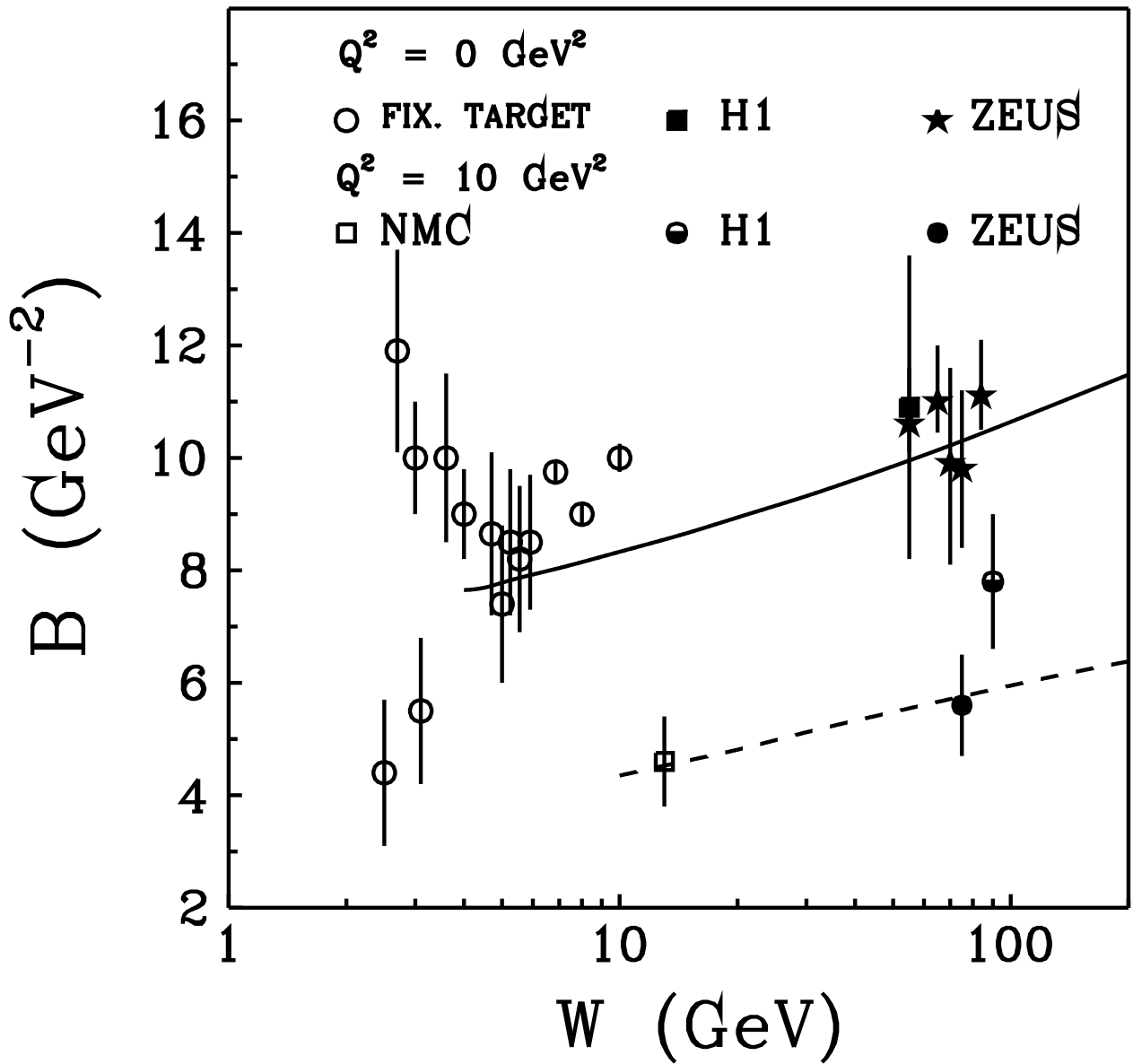
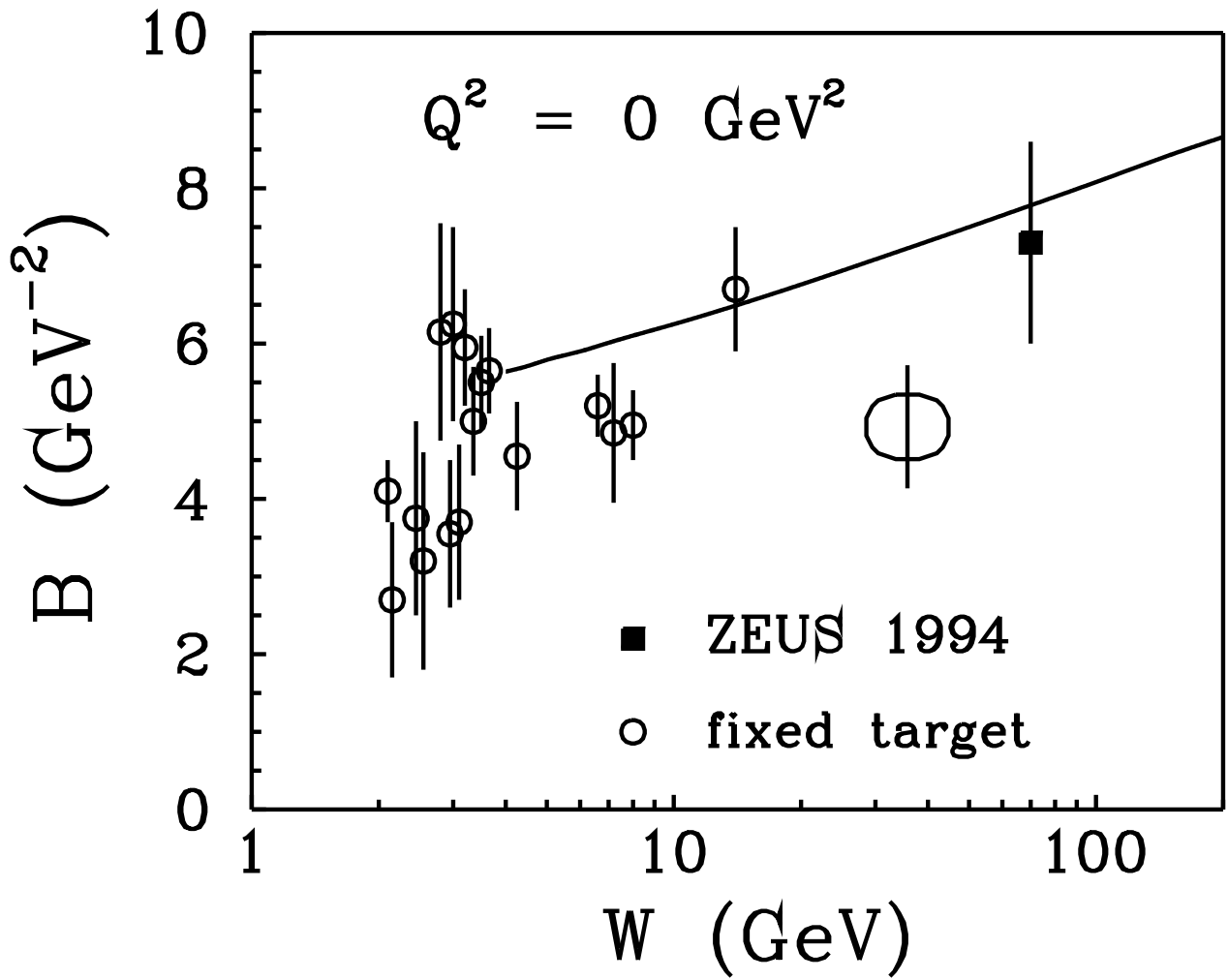
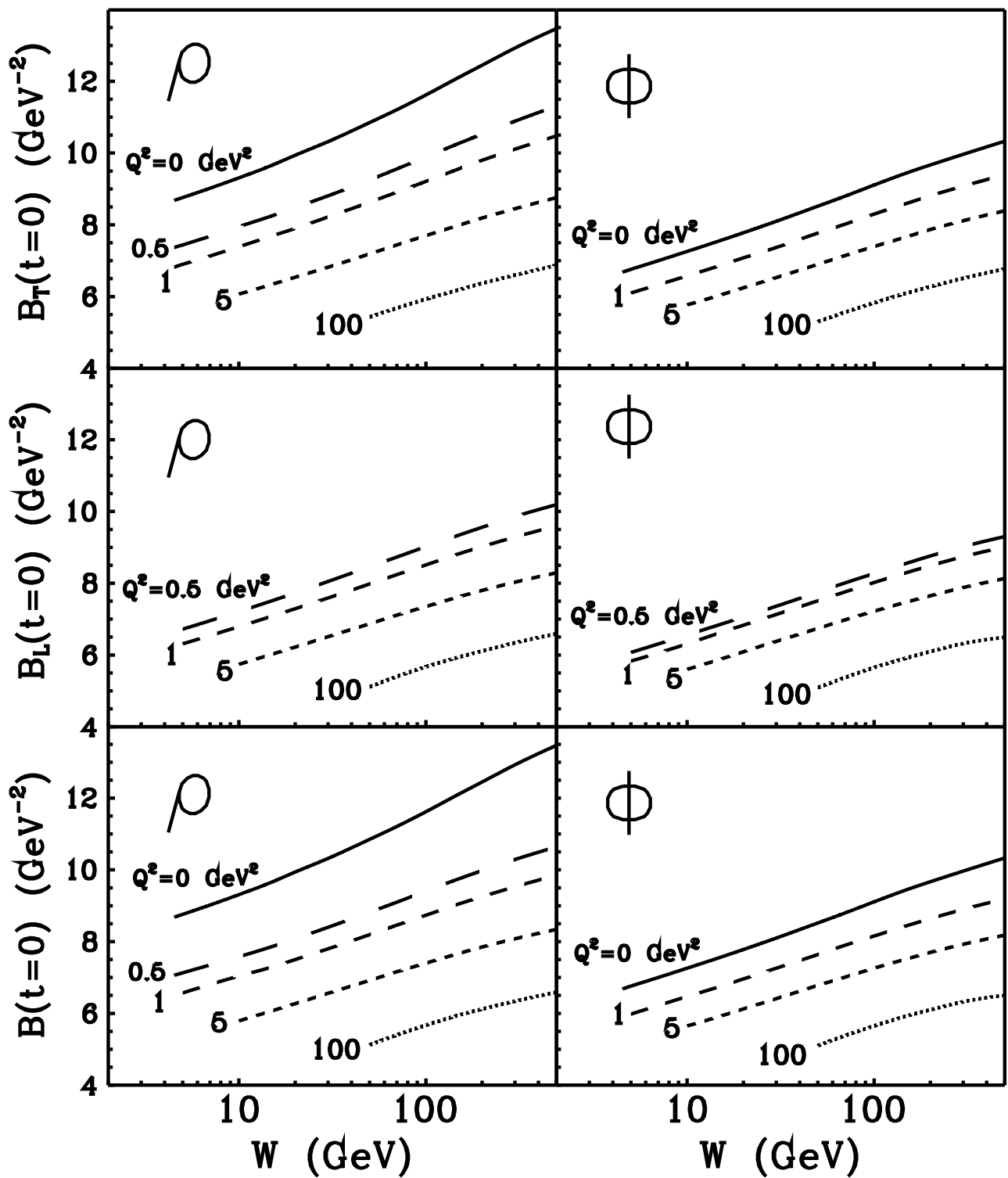
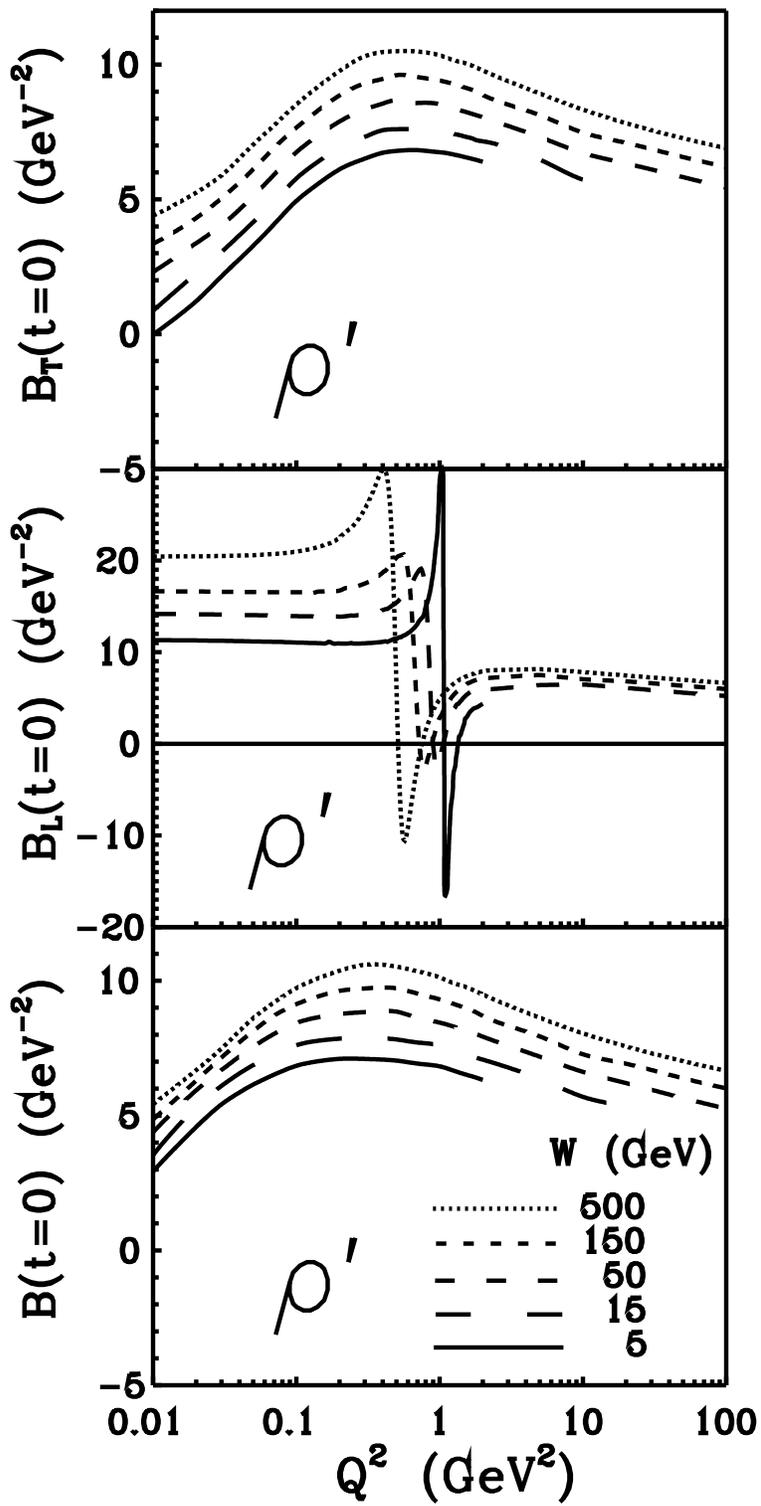
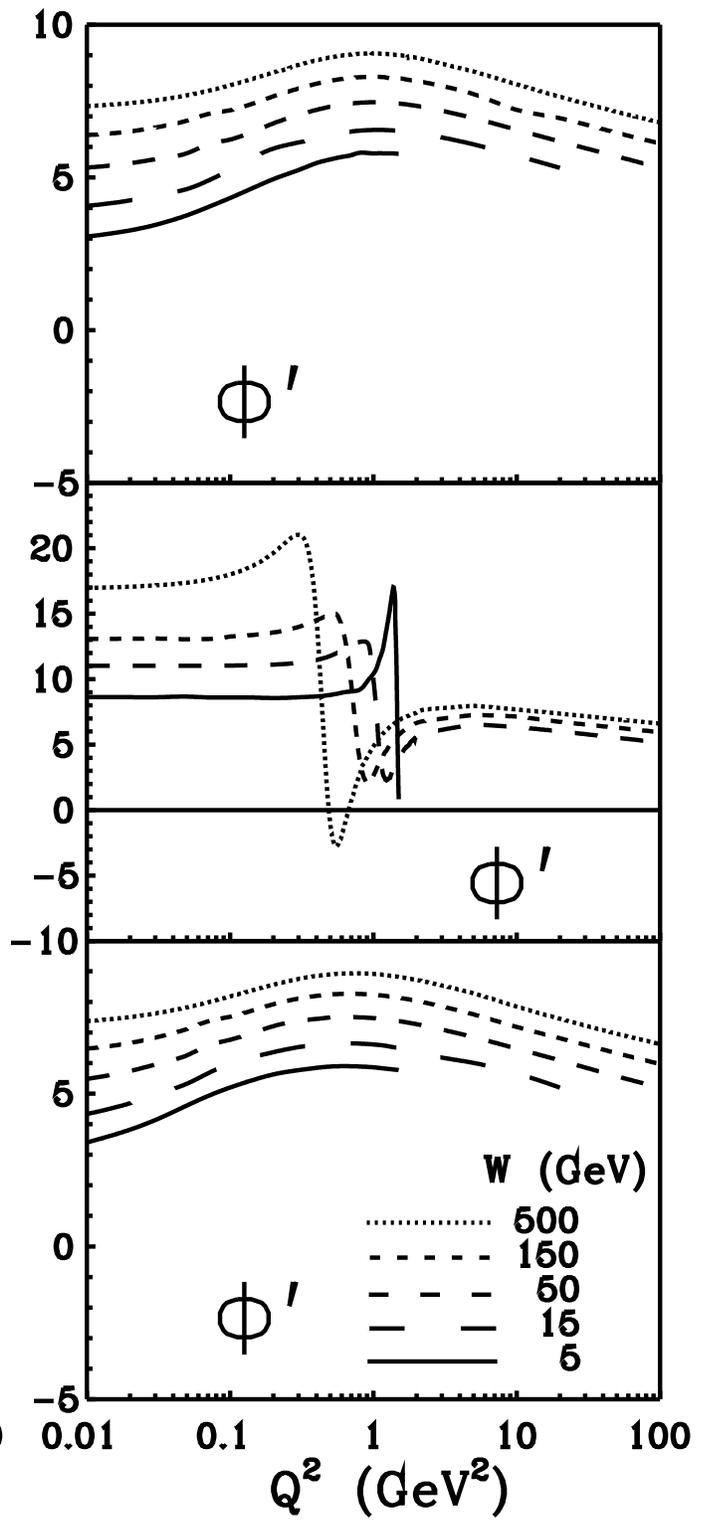
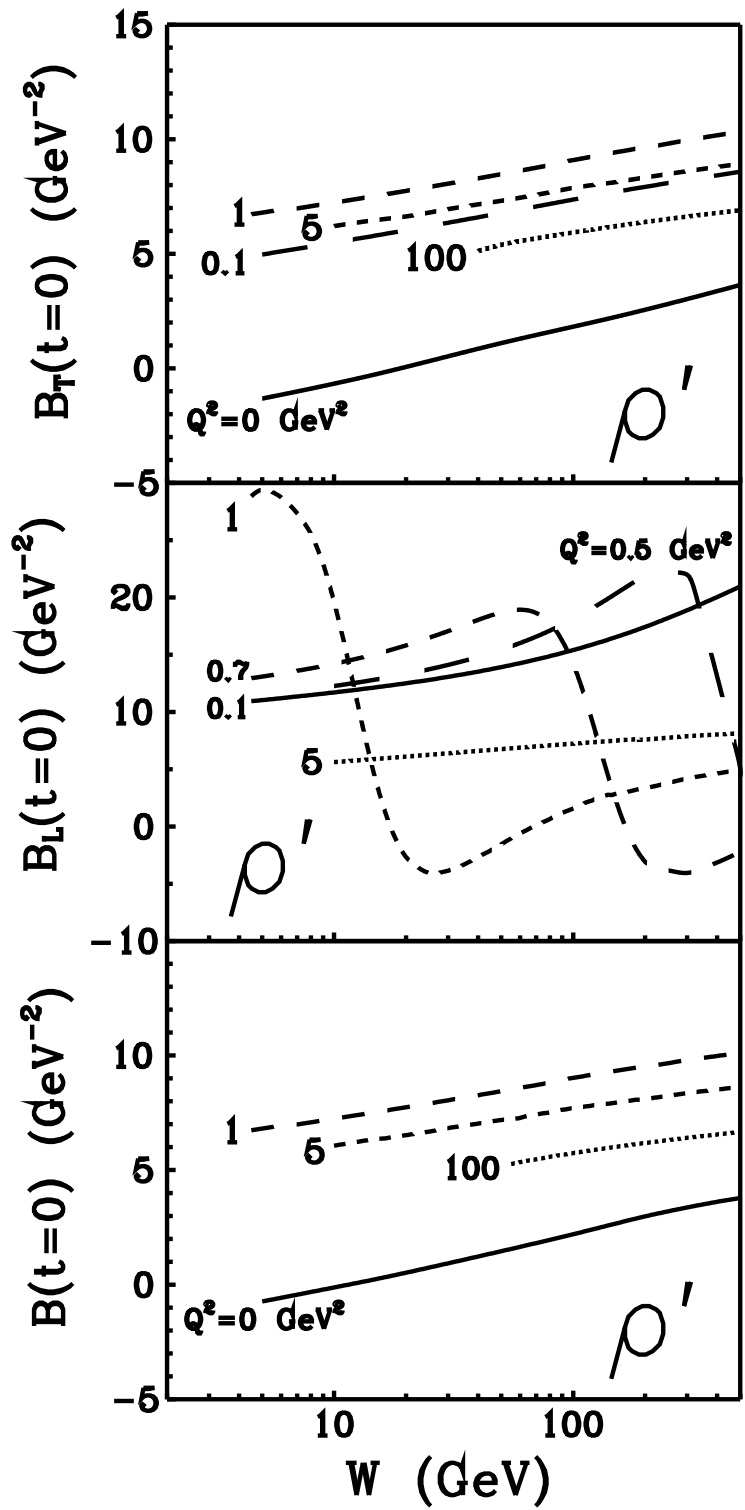
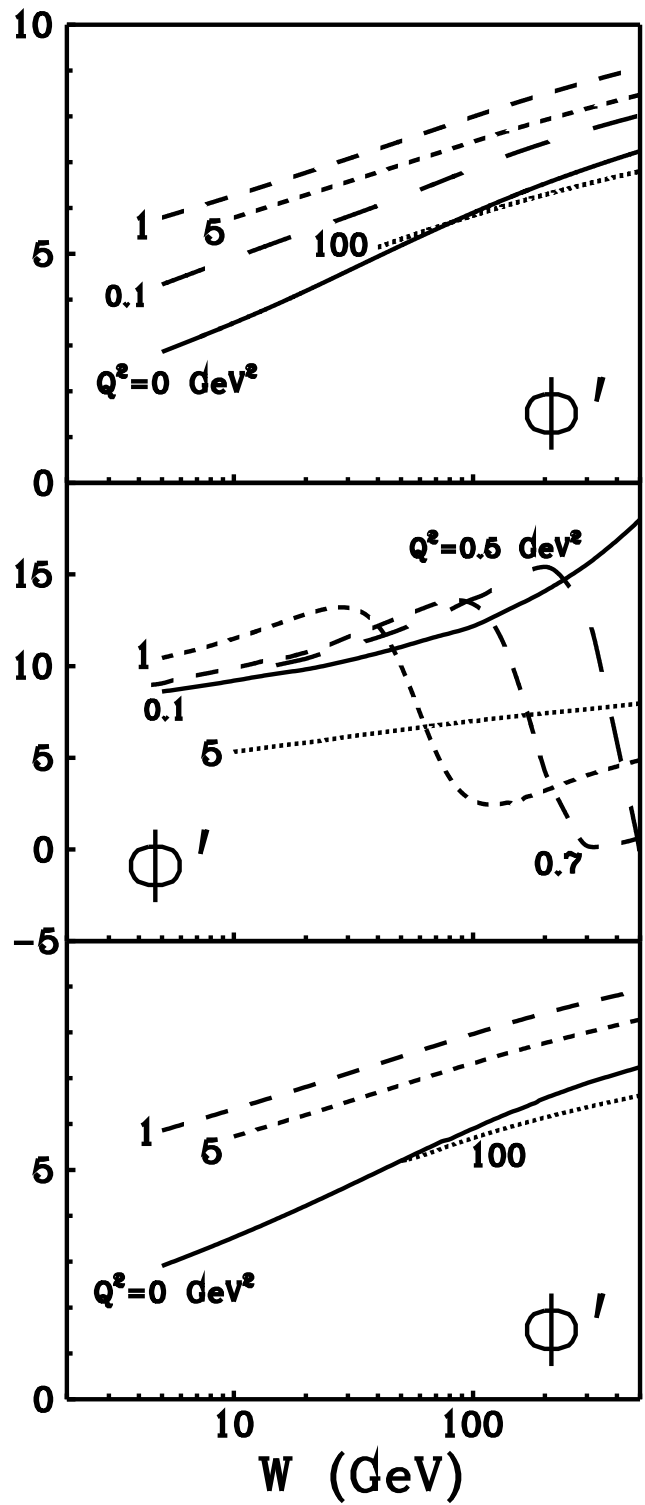
\end{document}